\documentstyle[12pt]{article}
\advance\topmargin by -5\baselineskip
\advance\textheight by 4\baselineskip
\begin{document}
\title{Quark fragmentation
functions in a diquark model for proton and $\Lambda$  hyperon production}


\author{\bf{Muhammad Nzar}   \\
Department of Physics    \\
Quaid-e-Azam University  \\
Islamabad, Pakistan. }
\date{}

\maketitle
\pagebreak

\begin{figure}
\begin{abstract}
A simple quark-diquark model for nucleon and $\Lambda$ structure is used to
calculate leading twist light-cone fragmentation functions for a quark to
inclusively decay into P or $\Lambda$. The parameters of the model are
determined by fitting to the known deep-inelastic structure
functions of the nucleon. When evolved from the initial to the final $Q^2$
scale, the calculated fragmentation functions are in remarkable agreement
(for $z>0.4 $) with those extracted from partially inclusive $ep$ and
$e^+ e^-$ experiments at high energies. Predictions are made, using no
additional parameters, for longitudinally and transversely polarized quarks
to fragment into p or $\Lambda$.
\end{abstract}
\end{figure}

\clearpage

Fragmentation of partons produced in a high-energy process into definite
hadronic states is a problem in QCD of considerable current
interest. Parton fragmentation, in a sense, is dual to the process of
uncovering hadron structure  using high momentum probes. This  duality has
recently been emphasized by Jaffe and Ji [1-2]. These authors have made a
detailed exploration of the spin, chirality, and twist structures of
fragmentation using the fact that parton   fragmentation functions can be
expressed in QCD as matrix elements of quark and gluon field operators
at light-cone separations. Indeed, a one-to-one correspondence between
fragmentation functions and parton distribution functions can be
established.   While this is a useful step towards understanding the
nature of the hadronization process, it certainly does not solve the
problem $\--$ the calculation of distribution functions belongs to the
realm of non-perturbative QCD and therefore can only be    modelled.
Fragmentation functions  are even harder to model theoretically, and
existing models using strings and shower algorithms etc.[3]  are complicated
and involve many parameters.

In this paper we explore an alternate, and much simpler, description for
the conversion of a fast quark into specific hadrons.
The starting point is an expression bilinear in quark field operators
which defines the unpolarized and polarized fragmentation functions in
terms of the light-cone momentum fraction $z = P^+/k^+$. (see fig-1),
\begin{equation}
4 \frac{{\hat{f}}_1(z)}{z} p^+ =  \int \frac{d\lambda}{2\pi} e^{- i
\lambda / z} Tr<0\mid \gamma^+ \psi(0) \mid PX><PX\mid \bar{\psi}(\lambda n)
\mid0>
\end{equation}
\begin{equation}
4 \frac{{\hat{g}}_1(z)}{z} p^+ S.n =  \int \frac{d\lambda}{2\pi} e^{- i
\lambda / z} Tr<0\mid \gamma^+ \gamma_5 \psi(0) \mid PS_\| X><P S_\| X\mid
\bar{\psi}(\lambda n)   \mid 0>
\end{equation}
and,
\begin{equation}
4 \frac{{\hat{h}}_1(z)}{z} S_\bot p^+  =  \int \frac{d\lambda}{2\pi} e^{- i
\lambda / z} Tr<0\mid \gamma^+ \gamma^\bot \gamma_5 \psi(0) \mid PS_\bot X><P
S_\bot X\mid
\bar{\psi}(\lambda n)   \mid 0>
\end{equation}
In the above, $p^\mu$ and $n^\mu$ are null vectors with $p^2 = n^2 = p^- = n^+
= 0$
and $p.n=1$. In terms of these, the  four momentum of the produced hadron,
whose
rest frame we shall take as our reference frame,  is $P^\mu = p^\mu +
\frac{1}{2} M^2 n^\mu $. The  spin  vector of the produced
spin-$\frac{1}{2}$ hadron is $S^\mu =  S.n p^\mu + S.p n^\mu +
M S^\mu_\bot $, and $\gamma^+ = 1/\sqrt{2} \left( \gamma^0 +
\gamma^3 \right)$. The gauge $n.A = A^+ = 0$ is used. A summation over
X is implicit and covers   all possible states which can be populated
by quark fragmentation. The quark operators are at equal $x_\bot$ and
$x^+$ but are separated by a variable light-cone distance $x^- = {\cal
\lambda} n^-$. To interpret $\hat{f}_1 (z)$ physically, define a
composite operator $C^\dagger (P)$ which creates a hadron of a specific
type and momentum from the vacuum, $\mid P > = C^\dagger (P) \mid 0 >$.
Using completeness of X, it is easy to show that,
\begin{equation}
 \hat{f}_1(z) =  \int [dk] \delta\left( \frac{k^+}{P^+} - \frac{1}{z}\right)
\frac{<k\mid C^\dagger(P) C(P) \mid k>}{<k\mid k >}
\end{equation}
where
\begin{equation}
[dk] = \frac{d^2 k_T dk^+}{(2\pi)^3 2 k^+}
\end{equation}
This shows that $\hat{f}_1(z)$ is the probability of finding a given hadron
with fixed z in a quark,  irrespective of the transverse quark momentum.

The matrix elements in eqs.1,2 need to be modelled as they cannot be
calculated ab-initio. Perhaps the simplest assumption is that the quark
fragments into a baryon and anti-diquark (fig.1b). This may be reasonable
provided that z is not too far from 1,
i.e. the produced hadron  carries away most of the momentum of the quark.
The amplitude for the process is,
\begin{equation}
<PS\mid \bar{\psi} \mid 0> = \bar{u}(PS)\Phi \frac{i}{k{\!\!\!/}-m}
\end{equation}
where $\Phi$ is the vertex that connects the quark with the two outgoing
particles.
$\Phi$ is a matrix in  Dirac space, and its most general form is rather
complicated since it involves several unknown form factors. Instead,
we shall be guided by the investigations of
Meyer and Mulders[4], and Melnitchouk, Schreiber, and Thomas[5], who
have calculated nucleon structure functions in the diquark model and
obtained rather good fits to deep inelastic scattering at all data
except small x. Following ref[4], we take
\begin{equation}
\Phi = \phi(P^2, p^{'2}, k^2)\,\,\sl{\cal I},
\end{equation}
where ${\cal I} $ is the unit matrix and only scalar diquarks are
included in the vertex. Inserting eqs.6,7 into eqs.1,2,3  and neglecting
the quark mass m in the numerator yields,
\begin{equation}
 \hat{f}_1(z) = \frac{1}{4(1-z)} \int_0^\infty \frac{d k^2_\bot}{(2\pi)^2}
\frac{M^2 + z^2 k^2_\bot}{(k^2 - m^2)^2} {\mid \! \phi \!\mid}^2
\end{equation}

\begin{equation}
 \hat{g}_1(z) = \frac{1}{8(1-z)} \int_0^\infty \frac{d k^2_\bot}{(2\pi)^2}
\frac{M^2 - z^2 k^2_\bot}{(k^2 - m^2)^2} {\mid \! \phi \!\mid}^2
\end{equation}
and,
\begin{equation}
\hat{h}_1(z) = \frac{M^2}{8(1-z)} \int_0^\infty \frac{d k^2_\bot}{(2\pi)^2}
\frac{1}{(k^2 - m^2)^2} {\mid \! \phi \!\mid}^2
\end{equation}

The fragmenting quark, which is highly off-shell, has its $k^2$ given by,
\begin{equation}
k^2 = \frac{M^2}{z} + \frac{m^2_d + zk_\bot^2}{1-z}.
\end{equation}
The integrals are logarithmically divergent and need to be damped by a
suitable form factor. Following ref.[5] we choose $\phi$ to be,
\begin{equation}
\phi(k^2) = N \frac{k^2 - m^2}{(k^2 - \Lambda^2)^2}.
\end{equation}
Here $ \Lambda $   is a mass parameter whose choice will be decided upon
later. N is a normalization constant which is fixed by the requirement
that the quark state be normalized covariantly,
$<k \mid k'> = (2 \pi)^3 2k^0 \delta^3(\vec{k}-\vec{k}')$.

Let us now concentrate on the spin-isospin structure of the produced
baryon. In the notation of Meyer and Mulders [4],   the SU(6) wavefunction
for a spin-up proton is:
\begin{equation}
\mid p \uparrow > = \frac{1}{\sqrt{2}} \stackrel{\rm \uparrow}{u} S^0_0 +
\frac{1}{\sqrt{18}} \stackrel{\rm \uparrow}{u} T^0_0
- \frac{1}{3} \stackrel{\rm \downarrow}{u} T^1_0
- \frac{1}{3} \stackrel{\rm \uparrow}{d} T^0_1
-\sqrt{ \frac{2}{9}} \stackrel{\rm \downarrow}{d} T^1_1.
\end{equation}
S and T are quark combinations representing scalar and vector diquark
respectively, e.g. $T^1_1 =  \stackrel{\rm \uparrow}{u}
\stackrel{\rm \uparrow}{u} $, etc. There is, of course, no explicit
coupling to vector diquarks in the vertex eq.7 although, at
the expense of some complication, it could be put in. We follow instead
the easy route suggested by the calculations of ref.[4] wherein it was
assumed that the S(x) and T(x) distribution functions are of identical
form but differ because the S and T  diquarks have somewhat different
 masses $m_s$ and $m_v$, as well as  mass parameters
$\Lambda_s$ and $\Lambda_v$. The $N-\Delta$ mass difference leads
to $m_v - m_s \approx 200 \, MeV$. Following a similar logic, define
$\hat{S}(z)$ and $\hat{T}(z)$ to be fragmentation functions where the
emitted anti-diquark is repectively S and T,
\begin{eqnarray}
\hat{S}(z) &=& \hat{f}_1(z) \,\,\,\, with \,\,\,\, m_d = m_s, \,\,
\Lambda = \Lambda_s       \nonumber \\
\hat{T}(z) &=& \hat{f}_1(z) \,\,\,\, with \,\,\,\, m_d = m_v, \,\,
\Lambda = \Lambda_v.
\end{eqnarray}
The polarized quantities $\Delta \hat{S}(z)$, $\Delta \hat{T}(z)$
are defined similarly with $\hat{f}_1(z)$ replaced by $\hat{g}_1(z)$
and $\Delta_1 \hat{S}(z)$, $\Delta_1 \hat{T}(z)$ are defined
with $\hat{f}_1(z)$ replaced by $\hat{h}_1(z)$
in the above. The u and d  fragmentation functions are readily seen to be,
\begin{eqnarray}
\hat{f}_{1u}^P(z) &=& \frac{1}{2} \hat{S}(z) + \frac{1}{6} \hat{T}(z)
\nonumber \\
\hat{f}_{1d}^P(z) &=& \frac{1}{3} \hat{T}(z)     \nonumber \\
\hat{g}_{1u}^P(z) &=& \frac{1}{2} \Delta \hat{S}(z) - \frac{1}{18} \Delta
\hat{T}(z)     \nonumber \\
\hat{g}_{1d}^P(z) &=& - \frac{1}{9} \Delta \hat{T}(z) \nonumber \\
\hat{h}_{1u}^{P}(z) &=& \frac{1}{2} \Delta_1 \hat{S}(z) - \frac{1}{18} \Delta_1
\hat{T}(z)   \nonumber \\
\hat{h}_{1d}^P(z) &=& - \frac{1}{9} \Delta_1 \hat{T}(z)
\end{eqnarray}

The same approch can be followed for the production of a $\Lambda$
hyperon from u, d, or s quarks. The hyperon wavefunction is,
\begin{equation}
\mid \Lambda \uparrow > = \frac{1}{\sqrt{12}} \left(
\stackrel{\rm \uparrow}{u} {\cal T}^0 - \stackrel{\rm \uparrow}{d}
{\cal T}^0 - \sqrt{2}  \stackrel{\rm \downarrow}{u} {\cal T}^1 +
\sqrt{2} \stackrel{\rm \downarrow}{d} {\cal T}^1
+ \stackrel{\rm \uparrow}{u} {\cal S}^0
+ \stackrel{\rm \uparrow}{d} {\cal S}^0
- 2 \stackrel{\rm \uparrow}{s} S^0   \right) .
\end{equation}
Here ${\cal S}$ and ${\cal T}$ are diquarks with one s and either a u
or d quark, while S is a ud system as before. Define new fragmentation
functions,
\begin{eqnarray}
\hat{\cal S}(z) &=& \hat{f}_1(z) \,\,\,\, with \,\,\,\, m_d = m_{\cal S},
\,\, \Lambda = \Lambda_{\cal S}       \nonumber \\
\hat{\cal T}(z) &=& \hat{f}_1(z) \,\,\,\, with \,\,\,\, m_d = m_{\cal T}
\!, \,\, \Lambda = \Lambda_{\cal T}       \nonumber  \\
\hat{S}(z) &=& \hat{f}_1(z) \,\,\,\, with \,\,\,\, m_d = m_s, \,\,
\Lambda = \Lambda_s      .
\end{eqnarray}
The quantities $\Delta \hat{\cal S}(z)$, $\Delta \hat{\cal T}(z)$,
$\Delta \hat{S}(z)$ are defined similarly with $\hat{f}_1(z)$ replaced
by $\hat{g}_1(z)$ and $\Delta_1 \hat{\cal S}(z)$, $\Delta_1 \hat{\cal T}(z)$,
$\Delta_1 \hat{S}(z)$ are defined  with $\hat{f}_1(z)$ replaced
by $\hat{h}_1(z)$. In terms of these, the fragmentation of u, d, and s
into $\Lambda$ is given by,
\begin{eqnarray}
\hat{f}_{1u}^{\Lambda}(z)\!\!\!\!\! &=& \!\!\!\!\! \hat{f}_{1d}^{\Lambda}(z)
\,\,=\,\, \frac{1}{4}
\hat{\cal T}(z) + \frac{1}{12}  \hat{\cal S}(z)    \nonumber \\
\hat{f}_{1s}^{\Lambda}(z) \!\!\!\!\! &=& \!\!\!\!\! \frac{1}{3} \hat{S}(z),
\end{eqnarray}
and for the polarized quantities,
\begin{eqnarray}
\hat{g}_{1u}^{\Lambda}(z) \!\!\!\!\! &=& \!\!\!\!\! \hat{g}_{1d}^{\Lambda}(z)
\,\, = \,\, \frac{1}{12}
\left( \Delta \hat{\cal S}(z) - \Delta \hat{\cal T}(z) \right)  \nonumber \\
\hat{g}_{1s}^{\Lambda}(z) \!\!\!\!\! &=&\!\!\!\!\!  \frac{1}{3} \Delta
\hat{S}(z) \nonumber \\
\hat{h}_{1u}^{\Lambda}(z) \!\!\!\!\! &=&\!\!\!\!\!  \frac{1}{12}
\left( \Delta_1 \hat{\cal S}(z) - \Delta_1 \hat{\cal T}(z) \right)  \nonumber
\\
\hat{h}_{1s}^{\Lambda}(z) \!\!\!\!\! &=&\!\!\!\!\!  \frac{1}{3} \Delta_1
\hat{S}(z)
\end{eqnarray}

We now turn to a discussion of numerical results and comparisons  with
fragmentation data, where available.  The diquark model for fragmentation
of u, d quarks to protons needs, as input, the scalar and vector diquark
masses $m_s$, $m_v$ as well as the cutoffs $\Lambda_s$, $\Lambda_v$.
Consistent  with the conclusions of  Melnitchouk et.al[5], we find that
a reasonable fit to nucleon structure functions can be achieved with the
choice $m_s \!\!=\!\! 900\,\, MeV$, $m_v \!\!=\!\! 1100\,\, MeV$
and $\Lambda_s\!\! =\!\! 840\,\, MeV$,
$\Lambda_v\!\! = \!\!925\,\, MeV$.  This fixes all the parameters needed for
the
model, and one may use it for calculating fragmentation rates into the
protons.  Fig.2 shows $\hat{f}_{1u}^{p}(z)$ and $\hat{f}_{1d}^{p}(z)$
calculated using eqs.8,12.   Since these are scale dependent quantities,
one must specify the scale as well. It appears reasonable to take the
initial scale to be $m_d + m_p$, where $m_p$ is the proton mass. This,
of course, is the minimum off-shell mass of the fragmenting quark.
Evolution of the fragmentation functions to the experimental  scale
can be performed exactly as for quark distribution functions. For example,
\begin{equation}
\mu \frac{\partial}{\partial \mu} f^P_i (z , \mu) = \sum_j \int^1_z
\frac{dy}{y} P_{i \rightarrow j} \left( \frac{z}{y} , \mu\right)
f^P_j (y,\mu),
\end{equation}
where $P_{i \rightarrow j} \left( x , \mu\right) $ is the Altarelli-Parisi
function for the splitting of the parton of type $i$ into a parton of type
$j$ with longitudinal momentum fraction x.
In principle, the sum in eq.20 extends over gluons and anti-quarks as well.
At large z this is hopefully small and  so we ignore this, including only
$P_{u \rightarrow u} \left( x , \mu\right) $ and
$P_{d \rightarrow d} \left( x , \mu\right) $ where,
\begin{equation}
P_{u \rightarrow u} \left( x , \mu\right) = \frac{2 \alpha_s (\mu)}{3\pi}
\left( \frac{1+x^2}{1-x} \right)_+  .
\end{equation}
The $ + $ function is defined in usual way,
\begin{equation}
f(x)_+ = f(x) - \delta (1-x) \int^1_0 dx' \, f(x') .
\end{equation}
The evolved fragmentation function $\hat{f}^P_u (z,Q^2)$ is compared
in fig.3 with the EMC data[6] extracted from $\mu - p$ and $\mu - D$
scattering. The agreement is fairly good  even down to rather small
values of z, where model has no reason  to be valid.  Without changing
parameters, predictions for $\hat{g}_1$ and $\hat{h}_1$ are shown in
figs.4,5    respectively.

The calculation of u, d, s quark fragmentation into $\Lambda$ hyperons
proceeds identically with the sole change of increasing the diquark mass
by $M_\Lambda - M_P = 186 MeV$ if the outgoing diquark contains a strange
quark.
The SU(6) wavefunctions then determine the relative magnitudes of
$\hat{f}_1$, $\hat{g}_1$, and $\hat{h}_1 $,  which are plotted respectively
in figs.6-8. That the produced $\Lambda$ carries the spin of the
fragmenting  s-quark is apparent from fig.7.
The transverse fragmentation function $\hat{h}_1$
 for $u, \, d, \, s \rightarrow \Lambda$, if measured, would be a good
tool for uncovering the transverse distribution function
$h_1(x)$ [1,10] of the proton. While no experimental spin data exists, data
from $e^+ e^-$
collisions at large cm energies can be compared with predictions of the
present model provided we assume that only u, d quarks lead to proton
production, and only u, d, s  quarks lead to $\Lambda$ production.
Fig.9 contains a comparison of theory to experiment.
 Again the evolution has been performed
to a final scale of $Q^2 = 80 \, GeV^2$.

To conclude, we have investigated a simple fragmentation model for a
quark to go into a $P, \, \, \Lambda$ and the appropriate anti-diquark, using
a vertex with a suitable form-factor determined by fitting to deep inelastic
data. The good agreement with
experiment$\-- $even down to rather small z values  where this agreement
is surely
fortuitous$\-- $suggests that one is perhaps using the right effective degrees
of freedom for $P, \,\, \Lambda $ production from quarks. At the same
time, it is obvious that one is far from a complete description
of fragmentation; the model gives zero chance for anti-proton
production from a quark. Neverthelesss, it has good predictive power,
 and we have calculated the spin-fragmentation  functions
$\hat{g}_1$ and $\hat{h}_1$ for $P, \,\, \Lambda$ production.
It will be interesting to see how well these are eventually
borne out by experiment.

\flushbottom
\centerline{\underline{\bf{Acknowledgement}}}
We thank Xiangdong Ji for a discussion and encouragement. M. Nzar thanks
Dr. Ahmad Ali for financial support during graduate study.
\pagebreak
\raggedbottom

\pagebreak
{\bf{\Large Figure Captions}}
\begin{enumerate}

\item The fragmentation of a quark with momentum k (fig.1a) into
a specific hadron (P, $\Lambda$, $\cdots$) is modelled (fig.1b) with a
quark-diquark-hadron vertex.

\item Fragmentation functions for u and d quarks to go
into a proton, calculated in the diquark model, evolved from the initial
scale $Q^2_0 = \left(m_d + m_p\right)^2$ to $Q^2 = 80 \,\, GeV^2$.

\item Evolved diquark model fragmentation function
$f^p_u (z,Q^2)$ compared against EMC[6] data extracted from
$\mu$ - P and $\mu$ - D scattering.

\item As in fig.2, except that the fragmentation
function $\hat{g}_1$ is plotted here.

\item As in fig.2, except that the fragmentation
function $\hat{h}_1$ is plotted here.

\item Diquark model calculation for u, d, s quarks to
fragment into $\Lambda$ hyperons. The initial scale  $Q^2_0 =
\left(m_d + m_\Lambda\right)^2$ and final scale $Q^2 = 80 \,\, GeV^2$
fragmentation functions are shown.

\item As in fig.6, except that the fragmentation
function $\hat{g}_1$ is plotted here.

\item As in fig.6, except that the fragmentation
function $\hat{h}_1$ is plotted here.

\item Production of protons and hyperons in $e^+e^- $
collisions at cm energy of about $30\,\, GeV$. The solid curves are predictions
of the diquark   fragmentation model, and the data is from TPC[7], HRS[8] and
MARK II[9] collaborations.
\end{enumerate}

Figures may be obtained by writing to\\
                    hafsa\%png-qau\%sdnpk@sdnhq.undp.org


\begin{thebibliography}{99}
\bibitem{kn:gnus} R. L. Jaffe and X. Ji, MIT CTP Preprint 2158(1993).
\bibitem{kn:gnus} X. Ji, MIT CTP Preprint 2219(1993).
\bibitem{kn:gnus}See, for example, ``Collider Physics",
V. D. Barger and R. J. N. Philips, Addison-Wesley, 1987.
\bibitem{kn:gnus}H. Meyer and P. J. Mulders, Nucl. Phys. {\bf A 528}, 589,
1991.
\bibitem{kn:gnus}W. Melnitchouk, A. W. Schreiber, and A. W. Thomas,
Phys. Rev. {\bf D 49}, 1183, 1994.
\bibitem{kn:gnus}M. Arneodo  et al, Nucl. Phys. {\bf B321}, 541, 1989.
\bibitem{kn:gnus} H. Aihara et. al (TPC) 1984, Phys. Rev. Lett. 53, 2378.
\bibitem{kn:gnus} M. Derrick et. al (HRS) 1987, Phys. Rev. D35, 2639.
\bibitem{kn:gnus} C. de la Vaissiere et. al (MARK II) 1985, Phys. Rev. Lett.
54, 2071.
\bibitem{kn:gnus}X. Artru and M. Mekhfi, Z. Phys. {\bf C45}, 669, 1990.


\end{thebibliography}
\end{document}